\documentclass{Interspeech}
\usepackage{amsmath,amssymb}
\usepackage{graphicx}
\usepackage{multirow}
\usepackage[table]{xcolor}
\usepackage{colortbl}
\usepackage{multirow}
\usepackage{booktabs}
\usepackage{pgf}

\interspeechcameraready

\title{Continuous Learning for Children's ASR: Overcoming Catastrophic Forgetting with Elastic Weight Consolidation and Synaptic Intelligence}

\author[affiliation={1}]{Edem}{Ahadzi}
\author[affiliation={1}]{Vishwanath}{Pratap Singh}
\author[affiliation={1}]{Tomi}{Kinnunen}
\author[affiliation={1}]{Ville}{Hautamaki}

\affiliation{}{University of Eastern Finland}{Finland}
\email{eahadzi@uef.fi, vsingh@uef.fi, tomi.kinnune@uef.fi, ville.hautamaki@uef.fi}

\usepackage{comment}

\begin{document}

\maketitle

\begin{abstract}
In this work, we present the first study addressing automatic speech recognition (ASR) for children in an online learning setting. This is particularly important for both child-centric applications and the privacy protection of minors, where training models with sequentially arriving data is critical. The conventional approach of model fine-tuning often suffers from catastrophic forgetting. To tackle this issue, we explore two established techniques: elastic weight consolidation (EWC) and synaptic intelligence (SI). Using a custom protocol on the MyST corpus, tailored to the online learning setting, we achieve relative word error rate (WER) reductions of 5.21 \% with EWC and 4.36 \% with SI, compared to the fine-tuning baseline.
\end{abstract}

\section{Introduction}

\emph{Automatic speech recognition} (ASR) is a mature technology that achieves high accuracy on varied and realistic speech data produced by adult speakers~\cite{xiong2017achievinghumanparityconversational, saon2017englishconversationaltelephonespeech}. Unfortunately, the overfocus on adult speech has hindered the development of ASR systems for children. While essential in applications such as educational tools, communication aids, and interactive learning environments, children ASR performance remains low~\cite{articleChildSpeech, attia2024kidwhisperbridgingperformancegap}. Children’s speech poses unique challenges due to rapid developmental changes and distinct acoustic and linguistic characteristics \cite{attia2024kidwhisperbridgingperformancegap, shivakumar2021endtoendneuralsystemsautomatic, articleChildren}. Anatomically, children’s vocal tracts are shorter compared to adults, resulting in higher pitch and different formant structures \cite{ref5, ref6}. Linguistically, underdeveloped pronunciation skills lead to inconsistent articulation \cite{analysis_icslp}. Although the usual approach in large-scale machine learning tasks, including ASR, is to boost performance by increasing the quantity and diversity of training data, data resources available for children remain relatively scarce \cite{singh2022spectralmodificationbaseddata}.

The above challenges are further exacerbated by
ASR training methods, which typically rely on \emph{static}, \emph{non-evolving} datasets where models are exposed to a fixed corpus all at once \cite{Kheddar_2024}. This is problematic for any application where data arrives sequentially, with its underlying properties drifting over time. In addition, specifically relevant to children, strict privacy regulations such as the \emph{General Data Protection Regulation} (GDPR) in the European Union (Article 5(1)(e)) and the \emph{Children’s Online Privacy Protection Act} (COPPA) in the United States (16 CFR §312.10) mandate deletion of minors’ data once the immediate processing purpose has been met. These requirements make it impractical to store and revisit earlier datasets to be used for subsequent training runs. These constraints motivate the need for approaches that can efficiently train models on new data as it becomes available.

The common approach is to fine-tune or re-train the ASR model as soon as new data is available. While this approach readily adapts to the latest information, especially retraining is computationally demanding. But even more critically, both retraining and finetuning lead easily to so-called \emph{catastrophic forgetting} \cite{Kirkpatrick_2017, FRENCH1999128, zenke2017continuallearningsynapticintelligence}, a phenomenon where the model focus on the new data, but loses the previously acquired knowledge \cite{vandeven2024continuallearningcatastrophicforgetting}. An ASR system that performed well earlier is at risk of being "downgraded" due to its over-emphasis on the most recent data. 

There are various approaches for mitigating catastrophic forgetting. These include data-based methods (which store or generate past examples to replay and preserve knowledge) \cite{lopezpaz2022gradientepisodicmemorycontinual, li2017learningforgetting, sun2019lamollanguagemodelinglifelong, vonoswald2022continuallearninghypernetworks, saha2021gradientprojectionmemorycontinual}, parameter-regularization approaches (which penalize large changes in model parameters critical for previous tasks) \cite{Kirkpatrick_2017, schwarz2018progresscompressscalable, zenke2017continuallearningsynapticintelligence, aljundi2018memoryawaresynapseslearning, ehret2021continuallearningrecurrentneural}, and dynamic architecture techniques (which expand or reorganize model components to accommodate new tasks) \cite{rusu2022progressiveneuralnetworks, fernando2017pathnetevolutionchannelsgradient, mallya2018packnetaddingmultipletasks}.

We focus 
on two methods, \emph{elastic weight consolidation} (EWC)~\cite{Kirkpatrick_2017} and \emph{synaptic intelligence} (SI) ~\cite{zenke2017continuallearningsynapticintelligence}. Both are computationally efficient, robust, and theoretically grounded (see Section \ref{sec:methods}) approaches for balancing the retention of past knowledge with the acquisition of new information. EWC has been particularly successful in alleviating catastrophic forgetting in ASR. Prior studies have applied it 
to domain adaptation for different accents or geographic regions \cite{ghorbani2019domainexpansiondnnbasedacoustic, Trinh_2022}, multilingual learning \cite{pham2024continuallylearningnewlanguages, Qian_2024}, and recognition of \emph{out-of-vocabulary} (OOV) words \cite{Qu_2023}. \textbf{These studies, however, are focused on adult speech, leaving the question of the utility of EWC and SI in children ASR tasks unanswered. Our study seeks to address this knowledge gap.} It is useful to keep in mind that children are not merely `scaled down adults'; their speech exhibits substantial variability not only between different speakers, but also within the same speaker over relatively short time-scale, caused by developmental changes. This necessitates further investigation and re-assessment of results obtained for adult speaker populations, important for practical children ASR applications. In our experiments, we simulate gradual drift in speaker population through a novel experimental pipeline (on the otherwise standard MyST corpus), detailed in Section \ref{sec:experimenta-setup}. Our experiments also address practical consideration related to which (and how many) ASR models one can retain during online learning. Before the experiments, however, we begin with self-contained description of the parameter regularization methodology.

\section{Methods}\label{sec:methods}

\subsection{Overview of Parameter-Based Continual Learning}

In continual learning, a model incrementally acquires knowledge across a series of tasks, with the aim of maintaining performance on previously learned tasks while integrating new information. A key challenge in this paradigm is catastrophic forgetting, where updates for new tasks overwrite important representations acquired earlier.

Various approaches have been proposed to address this problem, including parameter regularization. The idea 
is to penalize large changes to previous model parameters, ensuring that new knowledge can be integrated without 'erasing' essential information from earlier tasks. Both EWC and SI selectively penalize changes to parameters identified as critical for previous tasks. Their optimization objective has the form, 

\begin{equation}
\label{eq:cl_general}
\mathcal{L}(\theta) = \mathcal{L}_{\mathrm{new}}(\theta) + \beta \sum_{i} \omega_i \left(\theta_i - \theta_i^*\right)^2,
\end{equation} where $\mathcal{L}_{\mathrm{new}}(\theta)$ is the loss on the current (new) task, $\theta^*$ denotes parameters learned from previous tasks and $\beta$ controls the degree of regularization. In EWC, $\beta$ is set as $\lambda/2$, where $\lambda$ is the hyper-parameter controlling regularization strength, and can be interpreted within a Bayesian framework as imposing a Gaussian prior over the parameters. Accordingly, the quadratic penalty corresponds to the negative log-prior, and $\beta$ reflects the inverse of the prior variance \cite{Kirkpatrick_2017}. In contrast, SI adopts an empirically driven approach to regularization. Here, a global scaling factor $\alpha$ serves a similar regularization role to that in EWC but is determined based on empirical observations rather than rooted on probabilistic principles~\cite{zenke2017continuallearningsynapticintelligence}. In both cases, the importance of a parameter $\theta_i$ is quantified by $\omega_i$: for EWC, $\omega_i$ is expressed as $F_{ii}$, the diagonal of \emph{Fisher information matrix} (FIM) that captures the sensitivity of the loss to changes in $\theta_i$; for SI, it is denoted as $\Omega_i^{T-1}$, a cumulative measure reflecting the parameter's contribution to previously learned tasks. We discuss these techniques in more detail in Sections~\ref{sec:ewc} and~\ref{sec:si}.

\subsection{Elastic Weight Consolidation}
\label{sec:ewc}

EWC is a continual learning technique that imposes a quadratic penalty on parameter deviations from the parameter vector
\(\theta_{T}^{*}\) learned in the most recent, 
\(T^{\text{th}}\) task~\cite{Kirkpatrick_2017}. Rather than storing and revisiting data from every previous task, it uses a single regularization term that reflects the importance of parameters, measured through the empirical FIM \cite{karakida2019universalstatisticsfisherinformation}.

Specifically, whenever the model transitions to a new task, EWC computes a single regularization term based on the parameters \(\theta_{T}^{*}\) from the most recently completed task and penalize deviations from these parameters in the loss function. This formulation is obtained by approximating the posterior distribution \(p(\theta \mid D_{\text{old}})\) from the previous task as a Gaussian via a Laplace approximation---with mean \(\theta_{T}^{*}\) and diagonal precision given by the Fisher information \(F\)---which yields a quadratic penalty term that constrains important parameters \cite{Kirkpatrick_2017}. Consequently, the modified loss function for a new task becomes:
\begin{equation}
\mathcal{L}(\theta) = \mathcal{L}_{\mathrm{new}}(\theta) + \frac{\lambda}{2} \sum_{i} F_{ii}(\theta_{i} - \theta_{T,i}^*)^2,
\end{equation} where \(\theta_{T}^{*}\) are the parameters learned from the immediately
preceding task \(T\), \(\lambda\) is a hyper-parameter controlling the strength of the regularization, and \(F\) is a measure of parameter importance derived from the FIM.

The FIM quantifies the sensitivity of a model’s log-likelihood to its parameters, identifying which parameters were most important for the model’s performance on the previous task \cite{karakida2019universalstatisticsfisherinformation}. It is defined as:
\begin{equation}
F = \mathbb{E} \Bigl[ \nabla_{\theta} \log p(x, y; \theta) \nabla_{\theta} \log p(x, y; \theta)^{\top} \Bigr],
\end{equation}
where \(x\) represents the input data, \(y\) represents the corresponding target outputs and \(\theta\) is the set of model parameters.
In practice, we approximate this expectation by using the \textit{empirical} FIM computed over the training
samples \(\{x^{(i)}, y^{(i)}\}_{i=1}^{N}\):
\begin{equation}
F \approx \frac{1}{N} \sum_{i=1}^{N} \sum_{k=1}^{C} 
\nabla_{\theta} f_{\theta,k} (x^{(i)}) 
\nabla_{\theta} f_{\theta,k} (x^{(i)})^{\top},
\end{equation}
where \(f_{\theta,k}\) is the \(k\)-th output of the neural network. 
The empirical FIM converges to the
true FIM as \(N \to \infty\) \cite{karakida2019universalstatisticsfisherinformation}.
In 
ASR, 
the output typically 
corresponds to the predicted probability for the \(k\)-th token (character, phoneme, or word) in the output vocabulary at a given time step. \(C\) is the total number of output units, that is, the size of the output vocabulary, and \(N\) is the number of training samples used for this computation. 

\subsection{Synaptic Intelligence}
\label{sec:si}

Synaptic intelligence is another 
regularization technique. Unlike EWC, which is a \emph{second-order} method 
relying on FIM, SI uses 
a \emph{first-order} measure of parameter importance \cite{benzing2021unifyingregularisationmethodscontinual}. SI tracks how each parameter $\theta_i$ is updated over successive training iterations and how these updates contribute to reducing the training loss \cite{zenke2017continuallearningsynapticintelligence}. Parameters whose changes result in more substantial decreases in the loss are deemed more important for preserving previously acquired knowledge.

Specifically, when training on a given task $T$, SI keeps a running tally of how each parameter $\theta_i$ changes, as well as how such changes affect the overall loss. After each training step, SI updates an importance factor $\Omega_i$ for parameter $\theta_i$, reflecting its importance for 
maintaining performance on task $T$ \cite{zenke2017continuallearningsynapticintelligence}. Consider an infinitesimal change $\delta(t)$ to the parameter vector $\theta(t)$ at training iteration $t$. The respective change in the loss can then be approximated by
\begin{equation}
\label{eq:loss_change}
\mathcal{L}(\theta(t)+\delta(t))-\mathcal{L}(\theta(t))\approx\sum_{k}g_k(t)\delta_k(t),
\end{equation} where \( g_k(t) = \partial L/\partial \theta_k \) is the gradient of the loss with respect to 
$\theta_k$. In this expression, each individual parameter change \(\delta_k(t)=\theta'_k(t)\) contributes \(g_k(t)\theta'_k(t)\) to the total loss change. To capture the cumulative effect over the entire training trajectory for task \(T\), we compute the integral of these contributions along the path in parameter space~\cite{zenke2017continuallearningsynapticintelligence}. This corresponds to:
\begin{equation}
\label{eq:path_integral}
\int_{t_{T-1}}^{t_T} g(\theta(t)) \cdot \theta'(t) \, dt,
\end{equation}
which, since the gradient is a conservative field, equals the loss difference between end and start point: \(\mathcal{L}(\theta(t_T)) - \mathcal{L}(\theta(t_{T-1}))\). Decomposing ~\eqref{eq:path_integral} as a sum over the individual parameters gives:
\begin{equation}
\label{eq:decomposition}
\begin{split}
\int_{t_{T-1}}^{t_T} g(\theta(t)) \cdot \theta'(t) \, dt &= \sum_{k} \int_{t_{T-1}}^{t_T} g_k(\theta(t)) \, \theta'_k(t) \, dt \\
&\equiv -\sum_{k} \omega_k^T.
\end{split}
\end{equation} Here, \(\omega_k^T\) represents the total contribution of parameter \(\theta_k\) to reducing the loss during task \(T\); a larger \(\omega_k^T\) indicates that changes in \(\theta_k\) were more instrumental in decreasing the loss \cite{zenke2017continuallearningsynapticintelligence}.

Once training on task $T$ completes, these importance measures \((\omega_k^T)_{k=1}^N\) are consolidated into final per-parameter importance scores \((\Omega_k^T)_{k=1}^N\). A normalization term is 
applied to ensure stability, which involves the total parameter change 
\(\Delta_k^T\) and a small damping constant \(\xi\). The damping constant \(\xi\) is introduced to bound the expression in cases where \(\Delta_k^T \to 0\) \cite{zenke2017continuallearningsynapticintelligence}:
\begin{equation}
\Omega_k^T \;=\; \sum_{\tau < T} \frac{\omega_k^\tau}{(\Delta_k^\tau)^2 + \xi}. \quad 
\end{equation}
Here, $\Delta_k^\tau = \theta_k(t_\tau) \;-\; \theta_k(t_{\tau-1}).$ This accumulation process yields a single importance weight $\Omega_k^T$ for each parameter, capturing how critical $\theta_k$ was deemed for task $T$ \cite{zenke2017continuallearningsynapticintelligence}.
Finally, 
the 
loss function for the new task becomes
\begin{equation}
\mathcal{L}_T(\theta) = \mathcal{L}_{\text{new}}(\theta) + \alpha \sum_{k}\, \Omega_k^{T-1}\,\bigl(\theta_k - \theta_k^{T-1}\bigr)^2,
\end{equation}
where \(\mathcal{L}_{\text{new}}(\theta)\) is the standard loss for the new task (cross-entropy in our experiments), \(\theta^{T-1}\) denotes the parameter values after training on previous tasks, \(\Omega_k^{T-1}\) represents the accumulated importance of parameter \(\theta_k\) from earlier tasks, and \(\alpha\) is a hyper-parameter controlling the regularization strength. 
It penalizes significant deviations of parameters that were crucial for past tasks, thereby helping to mitigate catastrophic forgetting.

\section{Experimental Setup}\label{sec:experimenta-setup}

\subsection{Dataset}

We utilize the \emph{My science tutor} (MyST) corpus \cite{pradhan2023sciencetutormyst}---one of the most extensive collections of children’s conversational speech in English. It comprises approximately 473 hours of audio from 1,371 students in grades 3 to 5, collected over 10,496 virtual tutoring sessions. The MyST project uniquely assigned each student’s data to one of three partitions---training, development, or test---thereby preventing speaker overlap.
 
During data preprocessing, we used only transcribed utterances (a requirement for supervised ASR training) and excluded utterances shorter than 0.5 seconds (which contain minimal linguistic content), and those longer than 30 seconds (which exceed the input constraints of the Whisper model \cite{radford2022robustspeechrecognitionlargescale}). After filtering, the dataset consists of 71,939 utterances (145.54 hours) for training from 567 unique speakers, 11,592 utterances (23.09 hours) for development from 80 unique speakers, and 12,578 utterances (25.05 hours) for testing from 91 unique speakers.

To simulate sequential data acquisition, we organized the training data into ten sequential \emph{utterance batches}. Unlike the '\emph{batch}' in machine learning---a randomly sampled subset \emph{of the same dataset} used at a single training step---the utterance batch is intended to represent a distinct snapshot \emph{of an evolving dataset} (see Table~\ref{tab:utterance_batch_stats}). 

\begin{table}[ht]
\centering
\scriptsize
\caption{Summary of utterance batch statistics, including the number of new speakers introduced (\(S_i\)), the total number of speakers (\(N_i\)) in each utterance batch (UB), and the corresponding number of utterances with their recording hours.}
\label{tab:utterance_batch_stats}
\begin{tabular}{lccc}
\toprule
\textbf{UB} & \textbf{New Speakers (\(S_i\))} & \textbf{Total Speakers (\(N_i\))} & \textbf{Utterance (Hrs)} \\
\midrule
1  & 56 & 56 & 6216 (12.84) \\
2  & 28 & 56 & 6275 (11.73) \\
3  & 28 & 56 & 6272 (11.81) \\
4  & 28 & 56 & 6203 (12.43) \\
5  & 28 & 56 & 6201 (12.32) \\
6  & 30 & 58 & 6128 (11.99) \\
7  & 31 & 59 & 6199 (12.92) \\
8  & 33 & 61 & 6133 (12.79) \\
9  & 31 & 59 & 6270 (12.83) \\
10 & 39 & 67 & 6194 (12.62) \\
\bottomrule
\end{tabular}
\end{table}

\noindent The utterance batches contain a partially overlapping, but progressively changing, set of speakers motivated by the natural decay in speaker recurrence in a real-world application. We model the retention of speakers from previous batches to decay exponentially; specifically, the number of speakers retained from an earlier batch \(j\) in a later batch \(i\) (with \(i > j\)) is given by
\begin{equation}
R_{ij} = \left\lfloor S_j \left(\frac{1}{2}\right)^{i-j} \right\rfloor,
\end{equation}
where \(S_j\) is the number of new speakers introduced in batch \(j\) and \((1/2)^{\,i-j}\) represents the exponential decay factor. The total number of speakers in batch \(i\) is then
\begin{equation}
N_i = \sum_{j=1}^{i-1} \left\lfloor S_j \left(\frac{1}{2}\right)^{i-j} \right\rfloor + S_i.
\end{equation}

\subsection{Model}

We adopt OpenAI's \emph{Whisper small} model \cite{radford2022robustspeechrecognitionlargescale} for our continual learning framework. This modern, high-performance ASR model pre-trained on around 680,000 hours of audio (including numerous accents and speech styles) represents a suitable trade-off between accuracy and computational efficiency, both important considerations to our resource-constrained lifelong learning setting. Whisper uses Transformer-based encoder-decoder architecture~\cite{vaswani2023attentionneed} with multi-headed self-attention~\cite{vaswani2023attentionneed} to capture both short- and long-range speech dependencies.

\subsection{Experiments}
\label{sec:exp}
 
In our experiments, we vary both the parameter regularization methods and model selection strategies. The former consists of (1) \textbf{no continual learning}, a model fine tuned incrementally on new data without any mechanism to prevent forgetting; (2) \textbf{EWC} as detailed in Section \ref{sec:ewc}; and (3) \textbf{SI} as detailed in Section \ref{sec:si}. The latter concerns strategy used for caching best-performing models in the online ASR setting. (i) First, the \textbf{no selection (NS)} strategy continues training the most recent model without re-evaluation, entailing minimum overhead but possibly resulting in performance degradation if the updated model fails to improve WER. (ii) \textbf{rolling window of 3 (RW3)} maintains the latest three models, requiring moderate storage and periodic validation to identify the best among these. While providing safeguard against performance drop more effectively, it also limits how many past models can be revisited. Thus, the final (iii) \textbf{best-of-all-so-far (BoA)} strategy keeps track of every model generated thus far, picking the top performer at each step. Although this 
guarantees strong performance, it may be computationally and memory intensive if many models must be stored and evaluated over time.

All the models are trained for three epochs. For hyper-parameter tuning, we experimented with learning rates ranging from $1.25\times10^{-5}$ to $1\times10^{-10}$ and regularization strengths ($\lambda$ for EWC and $\alpha$ for SI) of 0.1 and 0.01. We 
select the model that corresponds to the epoch with lowest WER on the development set. As for performance evaluation, we report word error rate (WER) on the development and test sets, including its 95\% confidence interval (computed using a standard bootstrap approach with multiple resamplings of the test data) \cite{inproceedingsBootstrap}.

\section{Results and Discussion}

Table \ref{tab:results} presents the batch-wise WERs for both the development and test data. Notably, when sequential fine-tuning is performed without any continual learning techniques, the development and test WER begins at a comparable WER with other experiments. However, an upward drift is evident up until the fourth utterance batch, after which the WER stabilizes at a marginally higher level. This upward trend in WER suggests that, over time, the absence of mechanisms to protect important parameters can lead to small, but noticeable, increase in WER.

In contrast, sequential fine-tuning with EWC and SI (i.e., the 'no selection' strategy) displayed relatively stable performance over the ten utterance batches. The WER on the development set after fine-tuning with EWC decreases from 18.94\% after utterance batch 1 to 18.75\% by utterance batch 4 without large deviations, while its test set WER decreases from 21.42\% after batch 1 to 21.11\% by batch 4. Fine-tuning with SI follows a similar pattern, with the development set WER moving from 18.94\% at utterance batch 1 to 18.80\% by utterance batch 4, and the test set WER from 21.42\% to 21.30\% by utterance batch 4. These results indicate that both methods retained knowledge from earlier tasks while adapting to new data, demonstrating effective mitigation of catastrophic forgetting.

\begin{table}[ht]
\caption{Development and Test Word Error Rate (WER, \%) after training on 10 sequential batches. For each method, B$n$ is the WER after batch $n$. `NS' stands for No Selection, `RW3' stands for `Rolling Window of 3' and `BoA' for `Best-of-All-So-Far.'}
\label{tab:results}
\centering
\setlength{\tabcolsep}{2pt}
\scriptsize
\resizebox{\columnwidth}{!}{%
  \begin{tabular}{l ll c *{10}{c}}
    \toprule
    \textbf{\rotatebox{90}{}} & \textbf{Exp.} & \textbf{Method} & \textbf{Set} & \textbf{B1} & \textbf{B2} & \textbf{B3} & \textbf{B4} & \textbf{B5} & \textbf{B6} & \textbf{B7} & \textbf{B8} & \textbf{B9} & \textbf{B10} \\
    \midrule
    \multirow{6}{*}{\rotatebox{90}{NS}} 
      & \multirow{2}{*}{(1)} & \multirow{2}{*}{No CL} & Dev  & 18.94 & 18.68 & 18.77 & 19.19 & 19.18 & 19.17 & 19.18 & 19.17 & 19.17 & 19.17 \\
      &                     &                      & Test & 21.42 & 21.32 & 21.98 & 22.27 & 22.27 & 22.27 & 22.27 & 22.27 & 22.27 & 22.27 \\
    \cmidrule{4-14}
      & \multirow{2}{*}{(2)} & \multirow{2}{*}{EWC}   & Dev  & 18.94 & 18.57 & 18.72 & 18.75 & 18.75 & 18.74 & 18.75 & 18.75 & 18.74 & 18.75 \\
      &                     &                      & Test & 21.42 & 21.08 & 21.30 & 21.11 & 21.11 & 21.11 & 21.11 & 21.11 & 21.11 & 21.11 \\
    \cmidrule{4-14}
      & \multirow{2}{*}{(3)} & \multirow{2}{*}{SI}    & Dev  & 18.94 & 18.57 & 18.78 & 18.80 & 18.79 & 18.79 & 18.79 & 18.79 & 18.79 & 18.79 \\
      &                     &                      & Test & 21.42 & 21.22 & 21.29 & 21.30 & 21.30 & 21.30 & 21.30 & 21.30 & 21.30 & 21.30 \\
    \midrule\midrule
    \multirow{6}{*}{\rotatebox{90}{RW3}} 
      & \multirow{2}{*}{(4)} & \multirow{2}{*}{No CL} & Dev  & 18.94 & 18.68 & 18.77 & 19.98 & 21.53 & 18.89 & 18.88 & 18.89 & 18.89 & 18.89 \\
      &                     &                      & Test & 21.42 & 21.32 & 21.98 & 23.24 & 27.38 & 21.33 & 21.33 & 21.18 & 21.18 & 21.18 \\
    \cmidrule{4-14}
      & \multirow{2}{*}{(5)} & \multirow{2}{*}{EWC}   & Dev  & 18.94 & 18.57 & 18.72 & 19.79 & 21.07 & 18.73 & 18.74 & 18.74 & 18.74 & 18.74 \\
      &                     &                      & Test & 21.42 & 21.08 & 21.30 & 22.79 & 25.69 & 21.30 & 21.30 & 21.30 & 21.30 & 21.30 \\
    \cmidrule{4-14}
      & \multirow{2}{*}{(6)} & \multirow{2}{*}{SI}    & Dev  & 18.94 & 18.57 & 18.78 & 19.78 & 20.60 & 18.79 & 18.79 & 18.79 & 18.79 & 18.79 \\
      &                     &                      & Test & 21.42 & 21.22 & 21.29 & 23.05 & 25.61 & 21.29 & 21.29 & 21.29 & 21.29 & 21.29 \\
    \midrule\midrule
    \multirow{6}{*}{\rotatebox{90}{BoA}}
      & \multirow{2}{*}{(7)} & \multirow{2}{*}{No CL} & Dev  & 18.94 & 18.68 & 18.77 & 19.98 & 21.53 & 20.39 & 19.37 & 19.21 & 19.75 & 19.88 \\
      &                     &                      & Test & 21.42 & 21.32 & 21.98 & 23.24 & 27.38 & 23.78 & 22.38 & 23.38 & 21.90 & 22.77 \\
    \cmidrule{4-14}
      & \multirow{2}{*}{(8)} & \multirow{2}{*}{EWC}   & Dev  & 18.94 & 18.57 & 18.72 & 19.79 & 21.07 & 19.69 & 19.26 & 19.64 & 19.89 & 19.84 \\
      &                     &                      & Test & 21.42 & 21.08 & 21.30 & 22.79 & 25.69 & 22.55 & 22.18 & 23.03 & 22.82 & 22.30 \\
    \cmidrule{4-14}
       
      & \multirow{2}{*}{(9)} & \multirow{2}{*}{SI}    & Dev  & 18.94 & 18.57 & 18.78 & 19.78 & 20.60 & 19.17 & 19.14 & 19.73 & 19.82 & 19.62 \\
      &                     &                      & Test & 21.42 & 21.22 & 21.29 & 23.05 & 25.61 & 22.33 & 21.90 & 22.38 & 21.97 & 22.54 \\
    
    \bottomrule
  \end{tabular}%
}
\vspace{-2mm}
\end{table}

Introducing the two model selection strategies, RW3 and BoA, adds another layer of complexity. For both the EWC and SI enhanced approaches---where continual learning is integrated with model selection via RW3 and BoA---there is a sudden performance spike around utterance batches 4 and 5. For example, EWC with the RW3 selection strategy sees the development set WER jump to 21.07\% at utterance batch 5 from 19.79\% in the previous utterance batch. However, the validation step allows the system to revert to a better-performing model from either the rolling window or the pool of past models, allowing the WER to revert to around 18.7–18.8\% on the development set. Similarly, non-continual learning with RW3 experiences a sharp drop in performance at utterance batch 5 (from 23.23\% to 27.38\% on the test set) but recovers by selecting a more favorable model from the rolling window. While model selection helps prevent the system from persisting with a model that has undergone a detrimental update, an unexpected speaker shift can briefly raise WER when the system reverts to a model lacking the latest speaker data.

A notable trend across all approaches is the eventual plateau in WER. Regardless of regularization strategy or model selection, the models converge to a narrow performance band. This plateau 
indicates limitations related to the homogeneity of data across batches, where each batch closely resembles the previous one. Even though we introduced speaker scheduling to increase diversity, the MyST dataset itsself is drawn from a relatively narrow age group (grades 3–5). This contributes to the overall homogeneity of the data and effectively constrains performance to a tight range.

\begin{figure}[ht]
    \caption{WER with 95\% Confidence Intervals obtained via bootstrap resampling across batches for each method}
    \centering
    \includegraphics[width=\columnwidth]{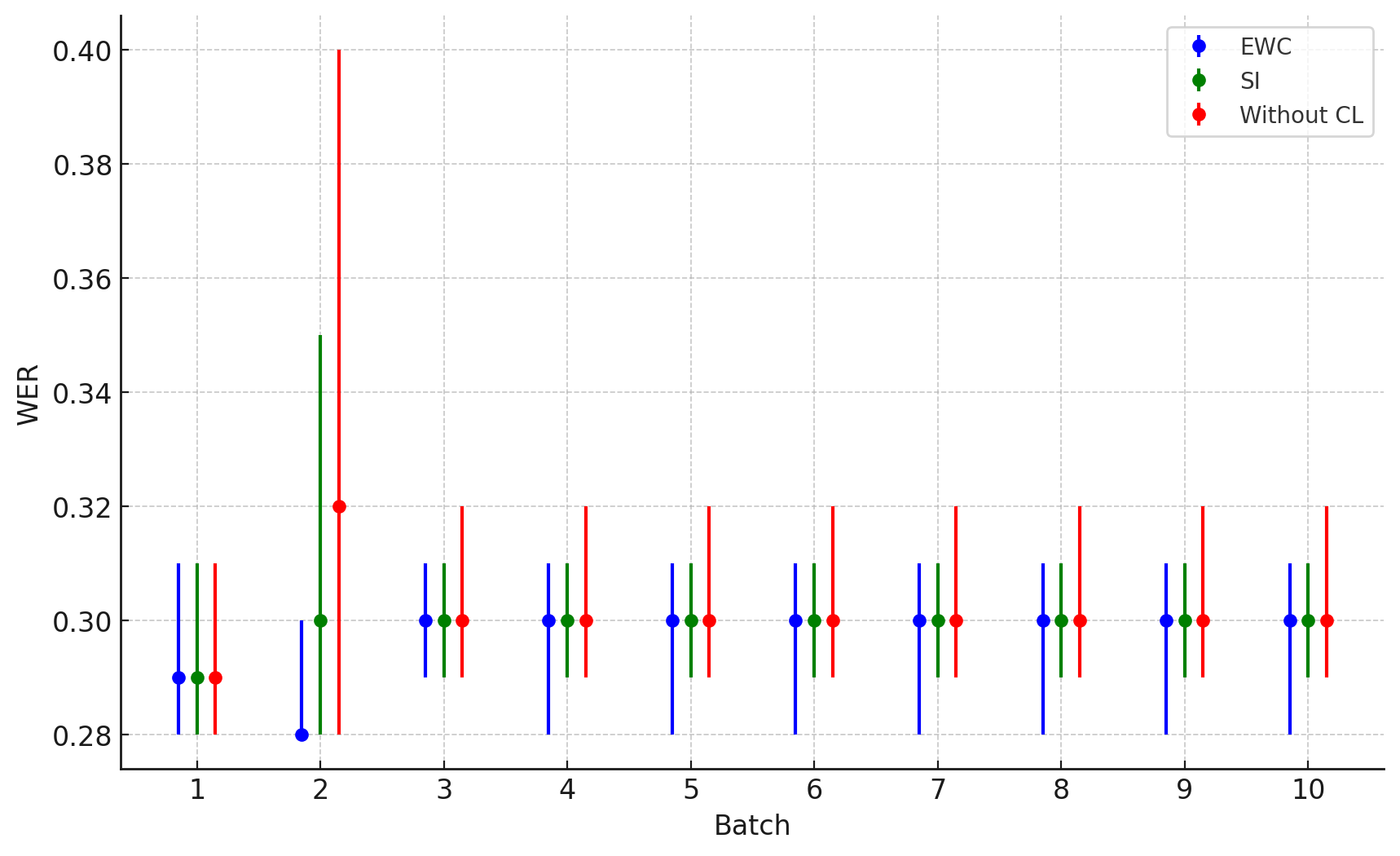} 
    \label{fig:wer_ci}
\end{figure}

A closer look at Figure \ref{fig:wer_ci}, which provides the average WER and corresponding 95\% confidence intervals, highlights how EWC and SI enhanced fine tuning differ statistically from non-continual learning. Both EWC and SI enhanced fine tuning and maintain relatively low average WERs with a narrow confidence intervals of 0.28-0.31 and 0.29-0.31, respectively. This narrower interval reflects the consistency observed in the batch-wise results. Meanwhile, fine-tuning without CL has an average WER ranging from about 0.29 to 0.32, with occasional 
confidence intervals reaching up to 0.40. This broader range suggests greater variability, aligning with the modest upward drift observed mid-training and the slightly higher final WERs in certain batches.

\section{Conclusion}

Our results on the MyST corpus demonstrate that the problem of catastrophic forgetting can be effectively mitigated using either of the two parameter regularization techniques,
EWC and SI. Both outperform standard fine-tuning in preserving earlier knowledge, while being capable of capturing new information. This dual capability has the potential of substantially enhancing online speech recognition for children, where data is continuously collected and speaker profiles evolve over time. While SI and EWC were found comparable in terms of WER, in practice SI is faster and is therefore recommended as the first choice for practitioners.

While our work provides a novel experimental validation of the efficacy of EWC and SI in children's online ASR tasks, we are constrained with the homogenous nature of the MyST corpus itself. Future work may consider e.g. simulated additive noise or reverberation to address robustness of EWC and SI to varied quality of data.

\section{Acknowledgement}
This study has been partially supported by the Academy of Finland (Decision
No. 349605, project "SPEECHFAKES")

\bibliographystyle{IEEEtran}
\bibliography{mybib}

\end{document}